\documentclass[a4paper,twocolumn,english,prb,showpacs,aps,superscriptaddress]{revtex4-1}
\usepackage[T1]{fontenc}
\usepackage[latin9]{inputenc}
\usepackage{amsmath}
\usepackage{amssymb}
\usepackage{graphicx}
\usepackage{esint}

\makeatletter

\pdfpageheight\paperheight
\pdfpagewidth\paperwidth

 \@ifundefined{textcolor}{}
 {%
   \definecolor{BLACK}{gray}{0}
   \definecolor{WHITE}{gray}{1}
   \definecolor{RED}{rgb}{1,0,0}
   \definecolor{GREEN}{rgb}{0,1,0}
   \definecolor{BLUE}{rgb}{0,0,1}
   \definecolor{CYAN}{cmyk}{1,0,0,0}
   \definecolor{MAGENTA}{cmyk}{0,1,0,0}
   \definecolor{YELLOW}{cmyk}{0,0,1,0}
 }

\usepackage{babel}

\makeatother

\usepackage{babel}
\begin{document}

\title{Numerical operator method for the real time dynamics of strongly-correlated
quantum impurity systems far from equilibrium}

\author{Pei Wang}

\affiliation{Department of Physics, Zhejiang University of Technology, Hangzhou
310023, China}

\email{wangpei@zjut.edu.cn}

\author{Guy Cohen}

\affiliation{Department of Chemistry, Columbia University, New York, New York
10027, U.S.A.}

\affiliation{Department of Physics, Columbia University, New York, New York 10027,
U.S.A.}

\author{Shaojun Xu}

\affiliation{School of Economics and Management, Zhejiang Sci-Tech University,
Hangzhou 310018, China}

\date{\today}
\begin{abstract}
We develop a method for studying the real time dynamics of Heisenberg
operators in strongly-interacting nonequilibrium quantum impurity
models. Our method is applicable to a wide range of interaction strengths
and to bias voltages beyond the linear response regime, works at zero
temperature, and overcomes the finite-size limitations faced by other
numerical methods. We compare our method with quantum Monte Carlo
simulations at a strong interaction strength, at which no analytical
method is applicable up to now. We find a very good coincidence of
the results at high bias voltage, and in the short time period at
low bias voltage. We discuss the possible reason of the deviation
in the long time period at low bias voltage. We also find a good coincidence
of our results with the perturbation results at weak interactions. 
\end{abstract}


\maketitle

\section{Introduction}

Understanding strongly-correlated open quantum impurity systems is
an important unsolved problem in condensed matter physics, and is
relevant to a wide variety of experimental fields. In cases ranging
from molecular electronics junctions,\cite{aviram_molecular_1974,jortner_molecular_1997,joachim_electronics_2000,metzger_quest_2001,heath_molecular_2003,joachim_molecular_2005}
low-dimensional mesoscopic systems\cite{datta_electronic_1997,goldhaber-gordon_kondo_1998,imry_introduction_2002}
and out-of-equilibrium correlated materials,\cite{asamitsu_current_1997,iwai_ultrafast_2003,liu_terahertz-field-induced_2012}
a need exists for reliable theoretical treatments which go beyond
linear response from equilibrium. The combination of correlated quantum
physics and the lack of a description in terms of equilibrium statistical
mechanics presents a major challenge in this regard, and while a variety
of powerful approximate schemes have been developed,\cite{datta_electronic_1995,plihal_transient_2005,haug_quantum_2008,karrasch_non-equilibrium_2010,pletyukhov_relaxation_2010,wang_flow_2010,swenson_application_2011,kennes_quench_2012,kennes_renormalization_2012,pletyukhov_nonequilibrium_2012,li_quasi-classical_2014}
the entire spectrum of interesting parameters is not reliably covered
by these methods. This makes numerically exact results highly desirable;
however, even when one is interested only in the properties of a stationary
state, the only recourse which does not involve approximations is
often to consider time propagation from some simple initial state
to the nonequilibrium steady state. Moreover, in some cases one can
observe a system's time-dependent response to a quench,\cite{perfetti_time_2006,kinoshita_quantum_2006,sadler_spontaneous_2006,bloch_quantum_2008}
and thus a theoretical treatment capable of following the system's
time evolution is needed.\cite{calabrese_time_2006,kollath_quench_2007,eckstein_thermalization_2009,schiro_time-dependent_2010,werner_weak-coupling_2010,kennes_quench_2012,cohen_greens_2014}

A great deal of progress has been made by considering the special
case of nonequilibrium quantum impurity models, where transport between
sets of infinite, trivial, noninteracting leads occurs through a finite,
nontrivial, interacting region.\cite{meir_landauer_1992,jauho_time-dependent_1994,datta_electronic_1997}
The nonequilibrium Anderson model, where the dot is modeled as a single
spin-degenerate electronic level with on-site Coulomb interaction,
exhibits a range of exotic phenomena related to Kondo physics\cite{bulla_numerical_2008}
and has drawn a particularly great deal of attention. While this model
and its extensions are still an infinite many-body problem and remain
under active research even at equilibrium,\cite{hewson_kondo_1993,bulla_numerical_2008,gull_continuous-time_2008,gull_continuous-time_2011,zgid_truncated_2012,ganahl_chebyshev_2014,lu_efficient_2014}
the local nature of the interactions in impurity models results in
major simplifications, and in recent years a great deal of progress
has been made in the development of numerically exact methods which
solve for transport properties in impurity models.\cite{anders_real-time_2005,anders_numerical_2008,muhlbacher_real-time_2008,weiss_iterative_2008,jin_exact_2008,schmidt_transient_2008,dias_da_silva_transport_2008,heidrich-meisner_real-time_2009,werner_diagrammatic_2009,wang_numerically_2009,segal_numerically_2010-1,eckel_comparative_2010,gull_numerically_2011,cohen_memory_2011,wang_numerically_2011,bedkihal_dynamics_2012,cohen_numerically_2013,cohen_generalized_2013,hartle_decoherence_2013,simine_path-integral_2013}
These controlled methods remain limited in their applicability; in
particular, quantum Monte Carlo (QMC) methods\cite{muhlbacher_real-time_2008,werner_diagrammatic_2009,gull_numerically_2011}
are generally slow to converge at low temperatures and long times,
while time-dependent density matrix renormalization group (tDMRG)\cite{dias_da_silva_transport_2008,heidrich-meisner_real-time_2009}
and numerical renormalization group (NRG)\cite{anders_numerical_2008,anders_steady-state_2008}
methods are effectively limited to systems where the leads can be
efficiently mapped onto reasonably short 1D chains with a small site
dimension, in such a way that entanglement along the chain remains
low. In recent years, impurity models have also been of interest as
auxiliary systems in the study of large or infinite interacting lattices,
by way of the DMFT approximation.\cite{georges_hubbard_1992,georges_dynamical_1996,kotliar_electronic_2006}

It is therefore necessary to continue exploring new numerical methods.
In this regard, the following observation is of some interest: when
studying real time dynamics, it is possible to solve the Schrödinger
equation and obtain the evolution of quantum states. However, quantum
states contain all the information about a quantum system, most of
which is redundant when one is only interested in very few or even
a single observable. An alternative route is to solve the Heisenberg
equation and obtain the evolution of a specific observable operators.
Solving the Heisenberg equation may allow for some simplification
or optimization. We note that similar ideas have been discussed in
the context of tDMRG, where matrix product operators can be more efficient
for describing dynamics than matrix product states.\cite{znidaric_many-body_2008,prosen_matrix_2009,hartmann_density_2009,banuls_matrix_2009,schollwock_density-matrix_2011,enss_light_2012}
The entanglement area laws which make DMRG perform so well for the
ground states of gapped systems, however, do not extend to nonequilibrium
situations and metallic systems. It is not therefore clear that tDMRG
should be an optimal scheme for such systems as impurity models. We
propose a different approach: to express the solution of the time-dependent
observable operators, we construct a set of basis operators, similar
to how one might choose basis vectors in the Hilbert space. As will
be shown, we can choose the basis operators so that the observable
that we are interested in is itself a basis operator at time zero.
The time-dependent observable operator then starts from a point on
an axis of the operator space and explores the other dimensions at
a finite rate, thus facilitating an efficient evaluation of the time
evolution in our basis. This allows us to work with an infinite model
and circumvent finite-size effects.

In this paper, we develop a numerical operator method (NOM) for nonequilibrium
quantum impurity models, based on previous work by one of the authors.\cite{wang_excitation_2013,wang_excitation_2013-1}
We apply the method to the real time transport dynamics of the Anderson
impurity model. Our method is in principle applicable to arbitrary
bias voltage and interaction strength. It describes infinite reservoirs
(which are difficult within tDMRG) at absolute zero temperature (which
is difficult for QMC), and can therefore be expected to be advantageous
in some regimes. We compare our results with perturbation theory at
weak interactions and with QMC at strong interactions. We find perfect
coincidence in most cases where it is expected, and discuss possible
reasons for deviations in problematic regimes.

The contents of the paper are arranged as follows. In sec.~\ref{sect:basis},
we introduce the model and the preliminary transformation of the Hamiltonian.
In sec.~\ref{sect:method}, we show the details of the NOM. In sec.~\ref{sect:result},
we show the validity and power of our method by giving some examples
and comparing our method with the others. At last, a concluding section
summarizes our results.

\section{The model and the transformation of the Hamiltonian}

\label{sect:basis}

\subsection{The Anderson impurity model}

The NOM is designed for solving the Heisenberg equation of motion.
In this paper, we discuss its application to the study of transport
through the nonequilibrium Anderson impurity model, an archetypal
model for the description of electron-electron interactions in quantum
junctions.\cite{meir_low-temperature_1993} The model involves an
impurity site coupled to two (``left'' and ``right'') electronic
reservoirs or leads: 
\begin{equation}
\hat{H}=\sum_{k,\alpha,\sigma}\epsilon_{k}\hat{c}_{k\alpha\sigma}^{\dagger}\hat{c}_{k\alpha\sigma}+\frac{g}{\sqrt{2}}\sum_{k,\alpha,\sigma}\left(\hat{c}_{k\alpha\sigma}^{\dagger}\hat{d}_{\sigma}+H.c.\right)+\hat{H}_{imp}.\label{andersonH}
\end{equation}
Here $\hat{d}_{\sigma}$ is an electronic annihilation operator at
the impurity, while $\hat{c}_{k\alpha\sigma}$ is an electronic annihilation
operator in the reservoirs. $\alpha\in\left\{ L,R\right\} $ denotes
the left and right reservoir, respectively, $\sigma\in\left\{ \uparrow,\downarrow\right\} $
denotes the spin and $k$ is an index corresponding to a reservoir
level with energy $\epsilon_{k}$. $g$ describes the coupling strength
between the impurity and the reservoirs (taken to be level-independent
here). $\hat{H}_{imp}$ is the local Hamiltonian at the impurity site,
and is expressed by 
\begin{equation}
\hat{H}_{imp}=\epsilon_{d}\sum_{\sigma}\hat{d}_{\sigma}^{\dagger}\hat{d}_{\sigma}+U\hat{d}_{\uparrow}^{\dagger}\hat{d}_{\uparrow}\hat{d}_{\downarrow}^{\dagger}\hat{d}_{\downarrow},
\end{equation}
where $\epsilon_{d}$ is the level energy and $U$ is the Coulomb
interaction. We concentrate on the particle-hole symmetric point $\epsilon_{d}=-U/2$
throughout the paper. We further define the impurity level broadening
$\Gamma=\rho\pi g^{2}$ ($\rho$ denoting the density of states of
the reservoir). As is customary in the field, $\Gamma$ will be used
as the unit of energy.

We assume an infinitely sharp cutoff in the reservoirs at a finite
bandwidth $D$. It is worth noting that our method can in principle
be used for an arbitrary frequency-dependent coupling $\Gamma\left(\omega\right)$.
We work at zero temperature, and take the chemical potentials of the
left and right reservoirs to be $\mu_{L}=V/2$ and $\mu_{R}=-V/2$,
respectively; $V$ is therefore a bias voltage across the junction.
At large $V$, the system is driven beyond the linear response regime
and can no longer be described well in equilibrium terms.

\subsection{The Wilson transformation}

We will not directly apply the NOM to the Hamiltonian Eq.~\eqref{andersonH},
but instead begin by discretizing the system and mapping it onto a
one-dimensional chain with only nearest-neighbor couplings by way
of a Wilson transformation. The reason for this is one of numerical
efficiency: the NOM works well when each creation and annihilation
operator appears in only a few terms of the Hamiltonian. In the transformed
Hamiltonian, this is true for operators either at the impurity site
or on the Wilson chain. However, in the original Hamiltonian Eq.~\eqref{andersonH},
the operator $\hat{d}_{\sigma}$ appears in an infinite number of
terms of the form $\left(\hat{c}_{k\alpha\sigma}^{\dagger}\hat{d}_{\sigma}+H.c.\right)$,
since the infinite reservoirs must be described by an infinite (or
at least large) number of $k$ indices. We note that the Wilson transformation,
which entails logarithmic discretization, is not a unique choice in
this regard: a more general Lanczos transformation allows for arbitrary
discretization schemes, and has been successfully employed in performing
similar mappings, for example in the context of recent DMRG\cite{ganahl_chebyshev_2014}
and configuration interaction\cite{lu_efficient_2014} solvers for
equilibrium impurity models. It should also be mentioned that the
Wilson transformation used was employed within the time-dependent
numerical renormalization group (tNRG) method to access the real time
dynamics of quantum impurity models coupled to both a single bath\cite{anders_real-time_2005}
and multiple baths.\cite{anders_steady-state_2008}

To proceed, it is useful to recombine the field operators in the two
reservoirs into pairs $\hat{c}_{k\pm\sigma}=\frac{1}{\sqrt{2}}\left(\hat{c}_{kL\sigma}\pm\hat{c}_{kR\sigma}\right)$,
where $\hat{c}_{k+\sigma}$ is called the symmetric operator and $\hat{c}_{k-\sigma}$
the antisymmetric operator. The Hamiltonian is then divided into symmetric
and antisymmetric parts: $\hat{H}=\hat{H}_{+}+\hat{H}_{-}$, where
\begin{equation}
\hat{H}_{-}\equiv\sum_{k}\epsilon_{k}\hat{c}_{k-\sigma}^{\dagger}\hat{c}_{k-\sigma}
\end{equation}
and 
\begin{equation}
\hat{H}_{+}\equiv\sum_{k}\epsilon_{k}\hat{c}_{k+\sigma}^{\dagger}\hat{c}_{k+\sigma}+g\sum_{k,\sigma}\left(\hat{c}_{k+\sigma}^{\dagger}\hat{d}_{\sigma}+H.c.\right)+\hat{H}_{imp}.
\end{equation}
The symmetric Hamiltonian describes an impurity coupled to a single
band, and can be transformed into a Wilson chain by a logarithmic
discretization of the band. Following ref.~\onlinecite{bulla_numerical_2008},
we then have 
\begin{equation}
\begin{split}\hat{H}_{+}= & \hat{H}_{imp}+\sqrt{\frac{\Gamma D}{\pi}}\sum_{\sigma}\left(\hat{d}_{0\sigma}^{\dagger}\hat{d}_{\sigma}+H.c.\right)\\
 & +\sum_{n=0}^{\infty}t_{n}\left(\hat{d}_{n\sigma}^{\dagger}\hat{d}_{n+1,\sigma}+H.c.\right),
\end{split}
\end{equation}
where $\hat{d}_{n\sigma}$ with $n=0,1,\cdots$ denotes the field
operator on the Wilson chain, $\Gamma$ the impurity level broadening,
$D$ the bandwidth of the reservoir, and $t_{n}$ the coupling between
neighboring sites on the chain. For constant $\Gamma$, an analytical
expression for the coupling strength can be obtained: 
\begin{equation}
t_{n}=\frac{D}{4}\frac{\left(1+\Lambda^{-1}\right)\left(1-\Lambda^{-n-1}\right)}{\sqrt{1-\Lambda^{-2n-1}}\sqrt{1-\Lambda^{-2n-3}}}\Lambda^{-n/2},
\end{equation}
where $\Lambda>1$ is a discretization parameter.

Finally, the full Hamiltonian consists of the combination of the antisymmetric
part and the Wilson chain, which are commutative with each other.
It can be expressed as 
\begin{equation}
\begin{split}\hat{H}= & \sum_{k}\epsilon_{k}\hat{c}_{k-\sigma}^{\dagger}\hat{c}_{k-\sigma}+\hat{H}_{imp}+\sqrt{\frac{\Gamma D}{\pi}}\sum_{\sigma}\left(\hat{d}_{0\sigma}^{\dagger}\hat{d}_{\sigma}+H.c.\right)\\
 & +\sum_{n=0}^{\infty}t_{n}\left(\hat{d}_{n\sigma}^{\dagger}\hat{d}_{n+1,\sigma}+H.c.\right).
\end{split}
\label{transformedH}
\end{equation}

In the limit $\Lambda\to1$, the transformed Hamiltonian Eq.~\eqref{transformedH}
is equivalent to the original Hamiltonian Eq.~\eqref{andersonH}.
We can therefore fully eliminate the error caused by the Wilson transformation
by taking the limit $\Lambda\to1$. It has been argued\cite{rosch_wilson_2012}
that the Wilson chain is not a thermal reservoir due to a finite heat
capacity, which scales as $1/\ln\Lambda$ in the limit $\Lambda\to1$,
such that the temperature of the Wilson chain is not fixed in a transport
setup. However, the dissipated energy in the setup within a finite
time is also finite, such that the chains simulate true reservoirs
for any given finite time if $\Lambda$ is close enough to $1$. Due
to the fact that the NOM operates on a set of truncated Heisenberg-picture
operators, the length of the Wilson chain does not significantly impact
the computational scaling and can essentially be taken to infinity
(see Fig.~\ref{fig:lambda} and the corresponding discussion there).
It can therefore be expected that it be valid to study the dynamics
up to some finite timescale even after a Wilson transformation with
$\Lambda>1$. In practice, for the timescales explored here, we find
that setting $\Lambda=1.2$ is sufficient for converging the discretization
error. We have verified that further reducing $\Lambda$ to $1.02$
does not significantly modify the results; we further note that within
standard tDMRG this is generally difficult to achieve and often values
of $\Lambda\simeq2$ are used.\cite{rosch_wilson_2012}

\section{The numerical operator method}

\label{sect:method}

\subsection{The current operator}

\label{subsect:currentop}

We study the current $I\left(t\right)\equiv\left\langle \hat{I}\left(t\right)\right\rangle $
through the impurity at a finite bias voltage $V$. The current operator
is given by 
\begin{equation}
\hat{I}\left(t\right)=-\frac{ig}{2}\sum_{k,\sigma}\left(\hat{d}_{\sigma}^{\dagger}\left(t\right)\hat{c}_{k-\sigma}\left(t\right)-H.c.\right),
\end{equation}
where we have employed the antisymmetric field operators $\hat{c}_{k-\sigma}\left(t\right)$
in order to express the difference between the currents as measured
in the left and right reservoirs. We assume that the reservoirs and
the impurity site are initially decoupled from each other. The two
reservoirs begin in their respective equilibrium states as determined
by the Fermi distribution $f_{\alpha}(\epsilon_{k})=\theta\left(\mu_{\alpha}-\epsilon_{k}\right)$,
while the impurity site is empty. We switch on the coupling $g$ at
$t=0$ and track the time evolution of the current. Since $\left[\hat{H}_{+},\hat{c}_{k-\sigma}\right]=0$,
it is straightforward to find that $\hat{c}_{k-\sigma}\left(t\right)=e^{-i\epsilon_{k}t}\hat{c}_{k-\sigma}$.
We therefore introduce a new field operator $\hat{c}_{-\sigma}=\frac{1}{\sqrt{\rho}}\sum_{k}e^{-i\epsilon_{k}t}\hat{c}_{k-\sigma}$,
and re-express the current as 
\begin{equation}
I\left(t\right)=2\sqrt{\frac{\Gamma}{\pi}}\mathrm{Im}\left\langle \hat{d}_{\sigma}^{\dagger}\left(t\right)\hat{c}_{-\sigma}\right\rangle .\label{currentexp}
\end{equation}
The problem is therefore reduced to the calculation of $\hat{d}_{\sigma}^{\dagger}\left(t\right)$,
which will be addressed by computational means in the following subsection.

\subsection{Iterative solution of the Heisenberg equation of motion}

To obtain the time dependence of $\hat{d}_{\sigma}^{\dagger}$, we
solve the Heisenberg equation of motion 
\begin{equation}
\frac{d\hat{d}_{\sigma}^{\dagger}\left(t\right)}{dt}=i\left[\hat{H},\hat{d}_{\sigma}^{\dagger}\left(t\right)\right].
\end{equation}
Since the antisymmetric Hamiltonian $\hat{H}_{-}$ commutes with $\hat{d}_{\sigma}^{\dagger}$,
this becomes 
\begin{equation}
\frac{d\hat{d}_{\sigma}^{\dagger}\left(t\right)}{dt}=i\left[\hat{H}_{+},\hat{d}_{\sigma}^{\dagger}\left(t\right)\right].\label{heisenbergHp}
\end{equation}
The symmetric Hamiltonian $\hat{H}_{+}$ describes a semi-infinite
Wilson chain with the impurity site as its first site. To simplify
the notation, we relabel the impurity site index as $-1$ such that
the impurity annihilation operator is $\hat{d}_{-1,\sigma}\equiv\hat{d}_{\sigma}$.
The symmetric Hamiltonian is then 
\begin{equation}
\begin{split}\hat{H}_{+}= & \epsilon_{d}\sum_{\sigma}\hat{d}_{-1,\sigma}^{\dagger}\hat{d}_{-1,\sigma}+U\hat{d}_{-1,\uparrow}^{\dagger}\hat{d}_{-1,\uparrow}\hat{d}_{-1,\downarrow}^{\dagger}\hat{d}_{-1,\downarrow}\\
 & +\sum_{\sigma,n=-1}^{\infty}t_{n}\left(\hat{d}_{n\sigma}^{\dagger}\hat{d}_{n+1,\sigma}+H.c.\right),
\end{split}
\label{symmetricHrelabel}
\end{equation}
where $t_{-1}\equiv\sqrt{\frac{\Gamma D}{\pi}}$.

In order to express $\hat{d}_{-1,\sigma}^{\dagger}\left(t\right)$,
we construct a set of basis operators. At each site $j=-1,0,\ldots$,
we choose some linearly independent set of sixteen local operators
generated by $\hat{d}_{j\sigma}^{\dagger}$ and $\hat{d}_{j\sigma}$
and including the unit operator. These are denoted by $\hat{\omega}_{j}^{i}$
for $i\in\left\{ 1,2,\ldots,16\right\} $, and every operator acting
only on site $j$ can be expressed as a linear combination of the
sixteen $\hat{\omega}_{j}^{i}$. We then propose that a basis operator
$\hat{O}_{\alpha}$ in the full symmetric subspace be expressed as
the product of on-site operators in an ascending order: 
\begin{equation}
\hat{O}_{\alpha}=\prod_{j=-1}^{\infty}\hat{\omega}_{j}^{\alpha_{j}}.
\end{equation}
Here $\alpha$ is an aggregate index representing a vector $\alpha_{j}$
with $j\in\left\{ -1,0,1,\ldots\right\} $, which identifies the basis
element in the full operator Hilbert space. The basis $\left\{ \hat{O}_{\alpha}\right\} $
is complete in the sense that any operator can be decomposed as a
linear combination of its members. Additionally, the coefficients
of this decomposition are unique.

In the basis just described the solution of the Heisenberg equation
Eq.~\eqref{heisenbergHp} can be written as 
\begin{equation}
\hat{d}_{\sigma}^{\dagger}\left(t\right)=\sum_{\alpha}a_{\alpha}\left(t\right)\hat{O}_{\alpha}.\label{obbasisexp}
\end{equation}
In tDMRG terms, one might say that our operator is represented by
a sum of terms of bond order 1, which is clearly very different from
a matrix product operator. To obtain the coefficient $a_{\alpha}\left(t\right)$
corresponding to each basis operator $\hat{O}_{\alpha}$, we derive
an iterative equation by propagating from time $t$ to time $t+\Delta t$
(where $\Delta t$ is some small time interval) by using the forward
Euler method: 
\begin{equation}
\hat{d}_{\sigma}^{\dagger}\left(t+\Delta t\right)=\hat{d}_{\sigma}^{\dagger}\left(t\right)+i\Delta t\left[\hat{H},\hat{d}_{\sigma}^{\dagger}\left(t\right)\right]+O\left(\Delta t^{2}\right).\label{obforward}
\end{equation}
We throw out terms of $O\left(\Delta t^{2}\right)$ and above, a valid
approximation in the limit $\Delta t\to0$. Next, substituting Eq.~\eqref{obbasisexp}
into Eq.~\eqref{obforward}, we obtain 
\begin{equation}
\hat{d}_{\sigma}^{\dagger}\left(t+\Delta t\right)=\sum_{\alpha}a_{\alpha}\left(t\right)\hat{O}_{\alpha}+i\Delta t\sum_{\alpha}a_{\alpha}\left(t\right)\left[\hat{H},\hat{O}_{\alpha}\right].\label{dtpdeltat}
\end{equation}

Calculation of the commutator $\left[\hat{H}_{+},\hat{O}_{\alpha}\right]$
is trivial and easily computerized. The important point is now that
at every stage of the computation, each $\hat{O}_{\alpha}$ appearing
in the expansion with a nonzero coefficient can be written as the
product of a finite number of (non-unit) local operators. Meanwhile,
each term in the Wilson Hamiltonian Eq.~\eqref{symmetricHrelabel}
involves at most four local operators which act either at the same
site or at two adjacent sites. Therefore, even though the Hamiltonian
$\hat{H}_{+}$ contains an infinite number of terms, the commutator
$\left[\hat{H}_{+},\hat{O}_{\alpha}\right]$ is always finite in length
for any finite $\hat{O}_{\alpha}$. In fact, $\left[\hat{H}_{+},\hat{O}_{\alpha}\right]$
generates only very few terms when $\hat{O}_{\alpha}$ is short, as
is the case when the propagation time $t$ is not too large. That
the commutator between the Hamiltonian and the basis operator contains
a finite number of terms is a necessary condition in order for the
NOM to be applicable. This condition can generally be satisfied for
lattice models with only short-ranged interactions, but will obviously
work best on low-dimensional lattices, which have a smaller coordination
number.

Let us write $\left[\hat{H},\hat{O}_{\alpha}\right]=\sum_{\alpha'}h_{\alpha,\alpha'}\hat{O}_{\alpha'}$
and substitute this into Eq.~\eqref{dtpdeltat}. We get 
\begin{equation}
\hat{d}_{\sigma}^{\dagger}\left(t+\Delta t\right)=\sum_{\alpha}a_{\alpha}\left(t\right)\hat{O}_{\alpha}+i\Delta t\sum_{\alpha,\alpha'}a_{\alpha}\left(t\right)h_{\alpha,\alpha'}\hat{O}_{\alpha'}.
\end{equation}
Noticing that $\hat{d}_{\sigma}^{\dagger}\left(t+\Delta t\right)=\sum_{\alpha}a_{\alpha}\left(t+\Delta t\right)\hat{O}_{\alpha}$
according to Eq.~\eqref{obbasisexp} and the expression of $\hat{d}_{\sigma}^{\dagger}\left(t+\Delta t\right)$
is unique due to our definition of basis operators, we finally obtain
the recurrence relation 
\begin{equation}
a_{\alpha}\left(t+\Delta t\right)=a_{\alpha}\left(t\right)+i\Delta t\sum_{\alpha'}h_{\alpha',\alpha}a_{\alpha'}\left(t\right).\label{coeffiterative}
\end{equation}
By using Eq.~\eqref{coeffiterative}, we can in principle obtain
the coefficients $a_{\alpha}\left(t\right)$ at arbitrary times by
an iterative procedure: we begin from an input $a_{\alpha}\left(t=0\right)$
which depends on the operator we wish to evaluate, and advance by
a sequence of time steps of size $\Delta t$.

We must store all nonzero coefficients $a_{\alpha}\left(t\right)$
along with their corresponding $\hat{O}_{\alpha}$ at time $t$ in
order to compute $a_{\alpha}(t+\Delta t)$. This demands that the
number of nonzero coefficients remain manageable. Fortunately, at
$t=0$ only a single nonzero coefficient is needed to express $\hat{d}_{\sigma}^{\dagger}$
(assuming that we choose $\hat{d}_{j,\sigma}^{\dagger}$ as one of
our $\hat{\omega}_{j}^{i}$). In a geometric picture, one could say
that our target $\hat{d}_{\sigma}^{\dagger}\left(t\right)$ begins
at $t=0$ exactly on an axis of the operator space at the initial
time, and the super operator $[\hat{H}_{+},]$ acting on $\hat{d}_{\sigma}^{\dagger}$
is quite inefficient at generating new nonzero dimensions. This property
guarantees that our algorithm be extremely fast at short times. This
property can only be taken care of within the Heisenberg picture:
in the Schrödinger picture, the operator $\hat{H}_{+}$ acting on
a unit basis vector (which is not an eigenstate of the Hamiltonian
or extremely local) will immediately generate an infinite number of
additional terms, stemming from the infinite number of terms in $\hat{H}_{+}$.
This difference between the two pictures is what makes the Heisenberg
picture a far more efficient one to study real time dynamics within
the NOM. This is understandable, since in the Heisenberg picture we
are restricting our focus to a single observable once a time, while
in the Schrödinger picture obtaining the quantum state is equivalent
to obtaining all possible observables and should in general be harder.

\subsection{The truncation scheme}

While the number of nonzero $a_{\alpha}\left(t\right)$ is always
finite, it also increases exponentially as the operator is propagated
to longer times. To keep the iterative process feasible, we must limit
the number of pairs $\left(a_{\alpha},\hat{O}_{\alpha}\right)$ that
are stored in memory. When the number of stored pairs exceeds a given
value (which we will denote by $M$), we perform a truncation and
throw out some number of the least important $\left(a_{\alpha},\hat{O}_{\alpha}\right)$.
The determination of the relative significance of the $\left(a_{\alpha},\hat{O}_{\alpha}\right)$
at each step, i.e., the truncation scheme, is critical: for our purposes,
a good algorithm is needed to allow the algorithm to accurately describe
$I\left(t\right)$ at long times.

To proceed in deriving an optimal truncation scheme for the current,
let us substitute the expansion for $\hat{d}_{\sigma}^{\dagger}\left(t\right)$
into Eq.~\eqref{currentexp} to obtain an expression for the current:
\begin{equation}
I\left(t\right)=2\sqrt{\frac{\Gamma}{\pi}}\sum_{\alpha}\mathrm{Im}\left\{ a_{\alpha}\left(t\right)\left\langle \hat{O}_{\alpha}\hat{c}_{-\sigma}\right\rangle \right\} .\label{currentbasisop}
\end{equation}
In considering this equation, an immediate and naive idea might be
to relate the magnitude $\left|\mathrm{Im}\left\{ a_{\alpha}\left(t\right)\langle\hat{O}_{\alpha}\hat{c}_{-\sigma}\rangle\right\} \right|$
to the significance of $\left(a_{\alpha},\hat{O}_{\alpha}\right)$.
However, this idea fails, since it leads to an underestimation of
the importance of $\left|a_{\alpha}\left(t\right)\right|$, which
has an \emph{inheritable} significance. To see this, one should consider
the fact that in the iterative relation Eq.~\eqref{coeffiterative},
a coefficient $a_{\alpha}\left(t\right)$ with a large magnitude also
has an important contribution to $a_{j'}\left(t+\Delta t\right)$.
On the other hand, a small coefficient $\left|a_{\alpha}\left(t\right)\right|\sim0$
can generally be thrown out, since its contribution to $a_{\alpha'}(t+\Delta t)$
is limited by 
\begin{equation}
\begin{split} & \left|\delta_{\alpha,\alpha'}a_{\alpha}\left(t\right)+i\Delta t\cdot h_{\alpha,\alpha'}a_{\alpha}\left(t\right)\right|\\
 & \leq\delta_{\alpha,\alpha'}\left|a_{\alpha}\left(t\right)\right|+\Delta t\left|h_{\alpha,\alpha'}\right|\left|a_{\alpha}\left(t\right)\right|,
\end{split}
\end{equation}
where the second term vanishes in the limit $\Delta t\to0$ (since
$\left|h_{\alpha,\alpha'}\right|$ is bounded). In the other words,
if $\left|a_{\alpha}\left(t\right)\right|$ is very small, throwing
out $\left(a_{\alpha},\hat{O}_{\alpha}\right)$ has no effect on the
current at the subsequent time. However, a coefficient with small
$\left|\mathrm{Im}\left\{ a_{\alpha}\left(t\right)\left\langle \hat{O}_{\alpha}\hat{c}_{-\sigma}\right\rangle \right\} \right|$
cannot be safely thrown out, since it may have a significant contribution
to $I\left(t+\Delta t\right)$ even though its contribution to $I\left(t\right)$
is essentially zero.

With this in mind, it is clear that using $\left|a_{\alpha}\left(t\right)\right|$
as our measure of significance is reasonable. However, it also has
the disadvantage of not being optimized specifically to the current.
Therefore, it may be better to give weight to the contributions of
both $\left|a_{\alpha}\left(t\right)\right|$ and $\left|\mathrm{Im}\left\{ a_{\alpha}\left(t\right)\left\langle \hat{O}_{\alpha}\hat{c}_{-\sigma}\right\rangle \right\} \right|$.
At short times, it is easy to see that the value of $\left|a_{\alpha}\left(t\right)\right|$
fluctuates strongly with different $\alpha$ (for instance, consider
$|a_{\alpha}(0)|$), such that $|a_{\alpha}\left(t\right)|$ is a
more important measure than $\left|\mathrm{Im}\left\{ a_{\alpha}\left(t\right)\langle\hat{O}_{\alpha}\hat{c}_{-\sigma}\rangle\right\} \right|$.
At long times, however, we find that the value of the contributing
$|a_{\alpha}\left(t\right)|$ at different $\alpha$ is of similar
size, a fact which may be understandable as a kind of thermalization
of $\hat{d}_{\sigma}^{\dagger}\left(t\right)$ in the operator space.
This leads to $\left|\mathrm{Im}\left\{ a_{\alpha}\left(t\right)\left\langle \hat{O}_{\alpha}\hat{c}_{-\sigma}\right\rangle \right\} \right|$
being more important at long times. In practice, we have used the
weight function 
\begin{equation}
W_{\alpha}\left(t\right)=\left|a_{\alpha}\left(t\right)\right|+\gamma e^{\beta t}\left|\mathrm{Im}\left\{ a_{\alpha}\left(t\right)\left\langle \hat{O}_{\alpha}\hat{c}_{-\sigma}\right\rangle \right\} \right|,\label{truncationscheme}
\end{equation}
where $\gamma,\beta>0$ are numerical parameters and the choice of
their value (which should affect the performance of the algorithm
but not the physical result) is decided empirically. After each time
step, we arrange the current set of stored operators and coefficient
$\left(a_{\alpha},\hat{O}_{\alpha}\right)$ in descending order according
to their respective weights $W_{\alpha}\left(t\right)$, and keep
the $M$ pairs $\left(a_{\alpha},\hat{O}_{\alpha}\right)$ with the
largest $W_{\alpha}$. The remaining pairs are discarded. Our experience
suggests that this truncation scheme performs far better at long times
than the more general alternative truncation scheme $W_{\alpha}\left(t\right)=\left|a_{\alpha}\left(t\right)\right|$,
which is not specifically tailored to the current.

\subsection{Evaluating the expectation value}

To obtain the current as expressed in Eq.~\eqref{currentbasisop},
we need to calculate the expectation value $\left\langle \hat{O}_{\alpha}\hat{c}_{-\sigma}\right\rangle $
with respect to the initial state. According to the definition of
basis operators, this requires evaluating expectation values of the
form 
\begin{equation}
\left\langle \prod_{j=-1}^{\infty}\hat{\omega}_{j}^{\alpha_{j}}\hat{c}_{-\sigma}\right\rangle .\label{wickexp}
\end{equation}
Since $\hat{\omega}_{j}^{\alpha_{j}}$ is a product of second quantization
operators, Eq.~\eqref{wickexp} can be expanded using Wick's theorem.
And we are able to calculate the contraction of an arbitrary pair
of field operators.

There are two kinds of nonzero contractions in Eq.~\eqref{wickexp},
which are $\left\langle \hat{d}_{n\sigma}^{\dagger}\hat{d}_{n'\sigma}\right\rangle $
and $\left\langle \hat{d}_{n\sigma}^{\dagger}\hat{c}_{-\sigma}\right\rangle $.
Using the initial condition of subsection~\ref{subsect:currentop},
a tedious but straightforward calculation gives 
\begin{equation}
\begin{split}\left\langle \hat{d}_{n\sigma}^{\dagger}\hat{d}_{n'\sigma}\right\rangle = & \sum_{m=0}^{\infty}\frac{u_{nm}u_{n'm}}{2d_{m}}\left\{ \int_{\frac{D}{2\Lambda^{m+1}}}^{\frac{D}{2\Lambda^{m}}}d\epsilon\left[f_{L}\left(\epsilon\right)+f_{R}\left(\epsilon\right)\right]\right.\\
 & \left.+\left(-1\right)^{n+n'}\int_{-\frac{D}{2\Lambda^{m}}}^{-\frac{D}{2\Lambda^{m+1}}}d\epsilon\left[f_{L}\left(\epsilon\right)+f_{R}\left(\epsilon\right)\right]\right\} 
\end{split}
\label{expectnnp}
\end{equation}
and 
\begin{eqnarray}
\langle\hat{d}_{n\sigma}^{\dagger}\hat{c}_{-\sigma}\rangle & = & \sum_{m=0}^{\infty}u_{nm}\sqrt{\frac{1}{d_{m}}}\label{eq:expectn}\\
 &  & \left[\int_{\frac{D}{2\Lambda^{m+1}}}^{\frac{D}{2\Lambda^{m}}}d\epsilon+(-1)^{n}\int_{-\frac{D}{2\Lambda^{m}}}^{-\frac{D}{2\Lambda^{m+1}}}d\epsilon\right]\nonumber \\
 &  & \thinspace\left[e^{-i\epsilon t}\frac{f_{L}\left(\epsilon\right)-f_{R}\left(\epsilon\right)}{2}\right],\nonumber 
\end{eqnarray}
where $f_{\alpha}\left(\epsilon\right)=\theta\left(\mu_{\alpha}-\epsilon\right)$
is the Fermi distribution, $d_{m}=\frac{D}{2}\left(\frac{1}{\Lambda^{m}}-\frac{1}{\Lambda^{m+1}}\right)$
the width of the discretized band, and $u_{nm}$ a set of orthogonal
coefficients in the Wilson transformation. $u_{nm}$ is generated
by the recurrence relation 
\begin{equation}
u_{n+1,m}=\frac{1}{t_{n}}\left(\frac{D\left(1+\Lambda\right)}{4\Lambda^{m+1}}u_{nm}-t_{n-1}u_{n-1,m}\right),
\end{equation}
with the initial conditions $u_{0m}=\frac{1}{\sqrt{2}}\sqrt{1-\Lambda^{-1}}\Lambda^{-m/2}$
and $u_{1m}=\frac{1}{\sqrt{2}}\sqrt{1-\Lambda^{-3}}\Lambda^{-3m/2}$.
A special case is $\left\langle \hat{d}_{n\sigma}^{\dagger}\hat{d}_{n'\sigma}\right\rangle =\frac{1}{2}\delta_{n,n'}$
for even $\left(n+n'\right)$.

There are four numerical (as opposed to physical) parameters in the
algorithm: the time interval $\Delta t$, the maximum number of stored
coefficients $M$, and the two parameters $\gamma$ and $\beta$ which
define the truncation weight function $W_{\alpha}$. Our algorithm
becomes numerically exact in the limit $M\to\infty$ and $\Delta t\to0$,
regardless of $\gamma$ and $\beta$. To obtain convergence, we start
from an initial guess $\left(\Delta t,M\right)$ for these values
and calculate the current $I\left(t\right){}_{\left(\Delta t,M\right)}$.
We then set $\Delta t\rightarrow\Delta t/2$ and $M\rightarrow2M$
and repeat the calculation to obtain $I\left(t\right){}_{\left(\Delta t/2,2M\right)}$.
The difference $\left|I\left(t\right){}_{\left(\Delta t,M\right)}-I\left(t\right){}_{\left(\Delta t/2,2M\right)}\right|$
provides us with an approximate estimate of the error, and we can
now iterate this procedure until the error is small enough for our
requirements, at which point we say that convergence is reached.

This concludes our discussion of the algorithm. In the next section,
we will present several examples and comparisons with other methods.

\section{Result and discussion}

\label{sect:result}

\subsection{Numerical parameters and convergence}

\begin{figure}
\includegraphics[width=1\linewidth]{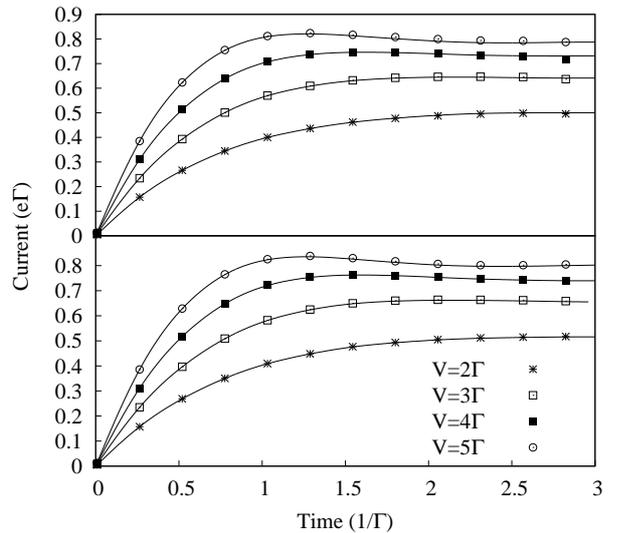} \protect\protect\caption{The time dependent current $I\left(t\right)$ computed at $\Lambda=1.02$
(represented by solid lines) and at $\Lambda=1.2$ (represented by
symbols) at several different bias voltages $V$ and two different
interaction strengths $U$. The results at $\Lambda=1.02$ coincide
with those at $\Lambda=1.2$, indicating that the latter is small
enough to attain convergence of the discretization of reservoirs at
these parameters and timescales. (Top panel) The interaction strength
is $U=\Gamma$. (Bottom panel) The interaction strength is $U=0.01\Gamma$.}

\label{fig:lambda} 
\end{figure}

We set the bandwidth of the reservoirs to $D=20\Gamma$ and calculate
$I\left(t\right)$ at the particle-hole symmetric point $\epsilon_{d}=-U/2$
for different interaction strengths $U$ and bias voltages $V$. The
choice of the truncation parameters $\gamma$ and $\beta$ is empirical.
We have tried different values in order to minimize the error of $I\left(t\right)$
at a given time for given $\Delta t$ and $M$, and found that $\gamma=2$
and $\beta=3$ is a good choice for a wide range of parameters. With
the truncation scheme we have proposed, computation time is efficiently
reduced to a manageable level: in practice we find that $M=40000$
and $\Delta t=0.008/\Gamma$ provide a good estimate of $I\left(t\right)$
for the parameters treated in this paper. In general, however, we
find that to obtain accurate results at longer times a larger $M$
is required, such that the computational scaling in the propagation
time $t$ is in practice substantially more than linear. In the general
case, this is of course a universal problem in all numerically exact
methods. We briefly mention that under certain conditions it can be
overcome by reduced dynamics techniques,\cite{cohen_memory_2011,cohen_numerically_2013,cohen_generalized_2013}
but this is beyond the scope of this paper.

An important problem that has been mentioned in sec.~\ref{sect:basis}
is that the Wilson chain is not a thermal reservoir at $\Lambda>1$.
We circumvent this issue by converging the data with the limit where
$\Lambda\to1$ for finite times. In Fig.~\ref{fig:lambda}, we present
the results at different $\Lambda$ for different interaction strengths
and bias voltages. We find that $\Lambda=1.2$ and $\Lambda=1.02$
give comparable results, indicating that the data is converged.

\subsection{Weak and intermediate interactions}

\begin{figure}
\includegraphics[width=1\linewidth]{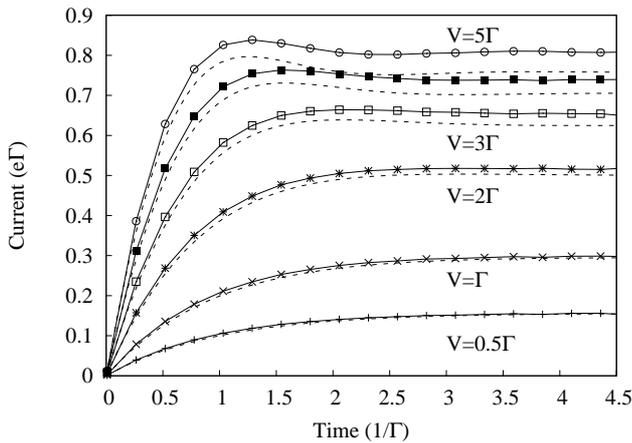} \protect\protect\caption{The current $I\left(t\right)$ at weak interaction strength $U=0.01\Gamma$
and for a range of bias voltages $\frac{V}{\Gamma}=\frac{1}{2},1,2,3,4,5$.
We compare results calculated with the NOM at finite bandwidth, $D=20$
(symbols connected by solid lines), with results calculated by second
order perturbation theory at infinite bandwidth, $D\rightarrow\infty$
(dashed lines).}

\label{fig:u001} 
\end{figure}

Having established that the NOM converges to a well-defined answer,
we now continue to argue that this answer is correct. This will be
done by performing a set comparisons with trustworthy analytical and
numerical results. We begin at the limit of weak interaction. For
noninteracting (quadratic) systems, the NOM has previously been discussed
in the literature and was shown to agree perfectly with the results
of exact diagonalization.\cite{wang_numerical_2014} We therefore
begin with weakly interacting systems at interaction strength $U=0.01\Gamma$,
where second order perturbation theory in $U$ may be expected to
work well.

In Fig.~\ref{fig:u001} we present a comparison between the NOM at
a bandwidth of $D=20\Gamma$ and perturbation theory. The perturbation
theory data comes from ref.~\onlinecite{wang_flow_2010}, where the
flow equation technique (which is equivalent at steady state to the
Keldysh technique of ref.~\onlinecite{fujii_perturbative_2003})
was applied to the Anderson impurity model at the wide band limit
$D\to\infty$. As expected, the two data sets converge at low voltages,
e.g., at $V=0.5\Gamma$ and $V=\Gamma$. We also find moderate deviation
of our results from the perturbation theory at high bias voltage,
where one might expect the bandwidth to be more important. In the
presence of interactions, the bandwidth affects the current even when
the transport window is narrower than the bandwidth ($V<D$): due
to inelastic scattering effects, levels in the bands are effectively
mixed.

\begin{figure}
\includegraphics[width=1\linewidth]{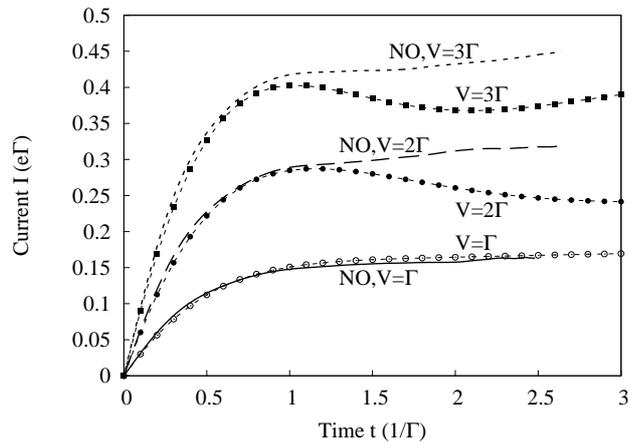} \protect\protect\caption{The current $I\left(t\right)$ at intermediate interaction strength
$U=4\Gamma$ for several different bias voltages $\frac{V}{\Gamma}=1,2,3$.
We compare results calculated with the NOM at finite bandwidth, $D=20$
(symbols connected by solid lines), with results calculated by second
order perturbation theory at infinite bandwidth, $D\rightarrow\infty$
(dashed lines).}

\label{fig:u4} 
\end{figure}

In Fig.~\ref{fig:u4}, we further compare our method with perturbation
theory at an intermediate interaction strength of $U=4\Gamma$ . Interestingly,
the numerical curves continue to fit well with the perturbative theory
at low bias voltages, but deviate from it qualitatively at high bias
voltage. This suggests that the second order perturbation theory still
works well at $U=4\Gamma$ and low voltage. At higher bias voltages
the deviation may be attributed to either the failure of the perturbative
approximation or the difference in bandwidth. The inelastic scattering
at $U=4\Gamma$ is stronger than that at $U=0.01\Gamma$, such that
the finite $D$ affects the current to a greater degree.

\subsection{Strong interactions}

\begin{figure}
\includegraphics[width=1\linewidth]{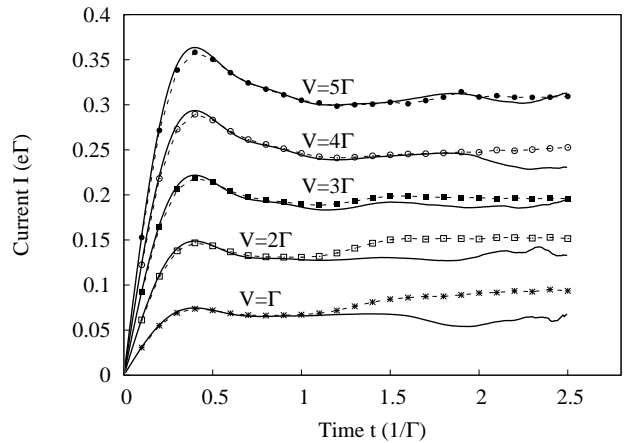} \protect\protect\caption{The current $I\left(t\right)$ at a large interaction strength $U=8\Gamma$
and several different bias voltages $\frac{V}{\Gamma}=1,2,3,4,5$,
as calculated by the NOM (solid lines) and the QMC method of ref.~\onlinecite{gull_numerically_2011}
(symbols connected by dashed lines).}

\label{fig:u8} 
\end{figure}

Finally, we study the current $I\left(t\right)$ within the strong
coupling regime by considering $U=8\Gamma$. In nonequilibrium---that
is, beyond linear response in the voltage---exact analytical results
are not available. Linear response is thought to be valid for voltages
approximately limited by the Kondo temperature $T_{K}\sim e^{-U/\Gamma}$,
and therefore quite small. We compare with numerically exact QMC results
at a low but finite temperature of $T=0.1\Gamma$, in comparison with
$T=0$ in the NOM. The short time behavior is accessible with continuous
time quantum Monte Carlo techniques,\cite{werner_diagrammatic_2009,gull_continuous-time_2011}
at least for temperatures which are not too low. Newer bold-line techniques\cite{gull_bold-line_2010,gull_numerically_2011}
allow access to longer times and lower temperatures. We have compared
the QMC results with similar at $T=0.2\Gamma$ (not shown) and verified
that within the timescales accessed here, the effect of the finite
temperature is small. Also, within QMC we take soft band edges with
a width of $0.1\Gamma$, as in ref.~\onlinecite{werner_diagrammatic_2009}.
The QMC simulations are otherwise performed at the same parameters
as the NOM.

The current $I\left(t\right)$ at different bias voltages up to $\Gamma t=2.5$,
as calculated by both methods, is displayed in Fig.~\ref{fig:u8}.
The results are consistent at the highest voltage $V=5\Gamma$. However,
even at $V=5\Gamma$, it is clear that the NOM gives a slightly larger
current around the first peak at $\Gamma t\sim0.5$, which can be
attributed to the difference in temperature and band shape. At lower
voltages the two methods exhibit good agreement at short times, but
deviate significantly from each other at longer times. Though steady
state is not reached here, NOM appears to predict a smaller steady
state current than QMC. We do not believe that this can be explained
by the difference in temperatures, since from the previously mentioned
check we find that reducing the temperature in QMC leads to the opposite
trend. QMC studies similarly suggest that the importance of the band
cutoff width $\nu$ is relatively unimportant at these parameters.
The rise in current at low voltages may be associated with the formation
of Kondo resonances at the chemical potentials,\cite{cohen_greens_2014}
and the failure of the NOM at these parameters indicates a problem
either with the Wilson mapping (which implies low resolution at high
energies) or with our truncation scheme. In either case, it requires
further investigation which will be left to future studies.

\section{Conclusions}

\label{sect:conclusions}

In summary, we have developed the numerical operator method, or NOM,
and applied it to the study of the real time dynamics of strongly-correlated
quantum impurity models in nonequilibrium. This is a notoriously difficult
problem to which many techniques have been applied. Our method is
distinguished by three important features, which we briefly outline
below.

First, in the mapping of the reservoirs onto 1D chains, any discretization
scheme is supported. This is similar to DMRG, but differs from NRG;
in QMC the issue of mapping onto a 1D chain need not arise. It also
allows us to efficiently take the limit $\Lambda\rightarrow1$ when
using the Wilson mapping.

Second, our method revolves around the solution of the Heisenberg
equation of motion. We carefully select a basis in the operator space
such that the superoperator $[\hat{H},\ldots]$ acting on our chosen
observable in this basis generates only a small number of terms. This
allows us to effectively set the length of the reservoir chains to
infinity, thus circumventing the finite-size scaling problems encountered
by other (non-QMC) numerical techniques.

Third, we provide a truncation scheme (see Eq.~\eqref{truncationscheme})
suited to the characteristics of the solution of the Heisenberg equation
and optimized for specific observables. Therefore, our algorithm is
extremely fast and accurate in the short time limit, and additionally
provides a controllable scheme for obtaining high quality approximations
of physical observables at longer times. The downside of this is that
our method only addresses a single observable per computation: to
calculate an additional observable, the entire time evolution process
must be repeated. However, in many cases, only one or very few observables
are of interest.

We note that while these features are also shared by tDMRG schemes
formulated for matrix product operators,\cite{schollwock_density-matrix_2011}
our truncation scheme is different and does not rely on the assumption
of low entanglement, which may not be appropriate for nonequilibrium
dynamics.

As an example, we apply our method to the real time dynamics of transport
through the nonequilibrium Anderson impurity model. We calculate the
time dependence of the current in a wide range of interaction strengths
and bias voltages going far beyond the linear response regime in both
quantities. We show that at small interaction strengths, our results
coincide with perturbation theory in the interaction. We further compare
our results with QMC data at a large interaction strength for which
no analytical method is known to be applicable, and find good agreement
as long as Kondo physics does not come into play.

We have therefore established the NOM as a reliable new formalism
for exploring nonequilibrium transport properties in the impurity
models over a wide range of parameters and at zero temperature. We
expect to generalize our method to more complicated quantum impurity
models in future and further analyze its advantages and limitations.
In particular, a comparison with tDMRG is in order, and the relative
merits of the NOM basis and the truncation scheme as opposed to those
of matrix product operator algorithms should be studied in detail.
Finally, modifications to DMRG which have allowed for the study of
finite temperatures\cite{feiguin_finite-temperature_2005,white_minimally_2009,stoudenmire_minimally_2010}
and open systems coupled to Markovian baths\cite{zwolak_mixed-state_2004,verstraete_matrix_2004,daley_atomic_2009}
should also be usable within the NOM. 
\begin{acknowledgments}
P. W. and S. X. were supported by the NSFC (Grants No. 11304280 and
No. 71103161). G. C. is grateful to the Yad Hanadiv-Rothschild Foundation
for the award of a Rothschild Postdoctoral Fellowship, and acknowledges
NSF CHE-1213247 and NSF DMR 1006282. QMC implementations were based
on the ALPS\cite{bauer_alps_2011} libraries. 
\end{acknowledgments}

 \bibliographystyle{apsrev4-1}
\bibliography{library}

\end{document}